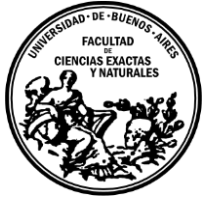

UNIVERSITY OF BUENOS AIRES

SCHOOL OF EXACT AND NATURAL SCIENCES

DEPARTMENT OF COMPUTING

# "TenisRank": A new ranking of tennis players based on PageRank

Thesis presented to qualify for the title of
Bachelor of Computer Science

## Alex Aronson

Director: Bachelor Ernesto Mislej

Co-Director: Dr. Flavia Bonomo

Buenos Aires, 2015

# "TENISRANK": A NEW RANKING OF TENNIS PLAYERS BASED ON PAGERANK

In the light of the need to achieve a ranking which is understood by all tennis supporters, the ATP ranking is exposed to constant complaints from players and at the same time exposes new players to be benefited with a good tournament to be able to start progressing in their careers. Moreover, the ATP ranking is not powerful enough to predict with certainty who will be the winner of a match if we are based solely on the positions.

In order to combat these problems, the idea of creating a new ranking that can indicate what are the real chances of victory of a player before the start of a new tournament arises. Based on the PageRank method, generated by Larry Page and Sergey Brin, we created a new ranking that specifically uses the characteristics of the tournament to generate data.

Based on a history of 40,000 matches, we intend to evaluate how the new method is performed as compared to other existing rankings in order to analyze if we really achieved an improved and real reflection.

Once we have obtained the ranking, we intend to evaluate, taking a sample game, the ranking of the players that dispute it and the characteristics of such game to be able to indicate the precise probability for the player with better ranking to win the game.

# ACKNOWLEDGMENTS

To Ernesto for having guided me, for his patience, his good predisposition and encouragement at each stage of this whole process.

To Flavia for the trust and for having helped me to find an idea that allows me to bring my two passions together: Sports and computing.

To my parents for each advice provided since primary school to this moment. For always worrying about my career evolution and insisting for me not to give up. For supporting me in each one of my decisions. I was given everything I needed to get here in the best way and I am infinitely grateful.

To Melu for accompanying me throughout my career unconditionally. For understanding my craziness attacks, my nerves before exams and helping me overcome them. For never letting me fall.

To my sisters for their every day support.

To my grandparents for always being around and happier for this moment than me.

To my friends for the thousand-cancelled hanging-outs, for the words of encouragement at all times.

To my fellow students for having helped me make the University a pleasant place.

To my co-workers for helping me reach the goal without putting obstacles in the way.

To Marcelo Albamonte for helping me visualize the number of things that can be done in sports based on what I learned at the University.

To Jorge for being my support in rough times and for accompanying me in good times.

To my University teachers, for all they thought me during these years.

*To my* zeide *Ernesto and my uncle Héctor. To my family and to Melu.*

# General index

# 1. RANKINGS IN GENERAL

## 1.1. Introduction to the Rankings

In our merit-based society, the concept of ranking is radical. In general, society consumes the best product, search engines show the most relevant document and people follow the statistics of their favorite team.

The massive consumption of rankings in various contexts made it necessary to develop many algorithms to achieve more efficient and reliable rankings.

As defined, a ranking is a classification from highest to lowest to set valuation criteria. In general, a ranking is materialized as a list that will set a relation between a set of elements joined in such list due to one feature in common.

Rankings are generally used in several areas to set some levels or determine them. This can appear in the musical environment as well as in the world of finance and business, for example, where we usually encounter different rankings of either most successful companies or best-selling products in the span of a month; richest men in the world, and the like.

Regarding fashion and beauty, these industries have rankings as well, which are prepared by specialized magazines in subjects such as, for example, best and worst dressed people.

Finally, and the focus of our study, this also happens in sports, where it is possible to meet with countless rankings such as the ranking of top scorers in soccer, best tennis players or the FIFA ranking that indicates which soccer teams are the best, among others.

## 1.2. Sports Rankings

Sports rankings are designed for several purposes. Some of them are commercial, for example, to ensure the presence of teams or players in some tournaments.

Perhaps equally important, both news media and economically interested observers in the game use rankings as a way of receiving advice and assessing who are the best players or teams today.

Moreover, players often mention as one of their biggest motivations the position in the ranking in which they and their team are qualified.

It can be argued that no ranking system accurately reflects the performance of each player, but it is rather a simplistic method to determine who is playing in the best way at any given time of the season.

Sports rankings also have a social function. In order to capture the largest number of people for them to consume professional tournaments, there may appear rankings that are simple to understand, so that all spectators can be attracted to the different matches of the tournaments.

However, in times when it is possible to have so much statistical information thanks to digital means which reflect data that were previously impossible to calculate easily, an opportunity appears to generate more accurate rankings to indicate who are the best players, but difficult for society to understand.

Another important use of the rankings is their predictive nature, where it can be determined who will be the winner of a match by looking at the positioning of the pertinent players. Many betting houses take these factors to place payment value for each play.

## 1.3.   Predictive Rankings v. Qualifying Rankings

In general, most ranking systems fail in one of two categories, either they are predictive, or they are qualifying.

The objective of the qualifying rankings is to indicate teams based on their participation in a certain competition throughout the season. Rankings allow to choose a champion or to use a reference team to indicate a set of teams that has ranked for its position in the ranking to participate in a particular tournament. The goal of a predictive ranking, however, is to provide the best prediction about the outcome of a match in which two players or teams will face each other.

An objective qualifying ranking should take as determining factors who was the winner or the score difference between players, or a combination of both. The use of a well-defined criterion allows teams or players to know exactly the consequences of winning, drawing, or losing a game. This is generally used in leaderboards for sports where each team or player receives a position, the lowest being the best ranked.

On the other hand, for the creation of predictive rankings as accurately as possible, it is allowed to include any useful additional information, such as: goals in favor of a team, matches won, scorers in the case of soccer or special features of certain tournaments and the backgrounds of each team at those tournaments or versus those rivals, among other aspects.

# 2. ATP RANKING

## 2.1.     Background

It is a universal goal of the players: to become the World's No. 1. Children dream of doing it, the effort is put into it. However, this is still one of the most elusive achievements in sports. In the 40 years of life of tennis rankings, only 25 players have reached the top of the (now) Emirates ATP Ranking, and only 16 finished the season as No. 1.

As stated in [1], since the beginning of the tennis Open Era in 1968, rankings were largely a subjective calculation generated by national associations, different circuits and specialized journalists who created their own lists.

Some players were on a roster of tennis players who could help sell tickets for the event, placing them as a priority over others on the tournament acceptance list. This caused great concern in those who did not have a big name and were at the verge of being excluded from events.

In August 1972, it became apparent that the newly created Association of Tennis Professionals would be necessary to establish a single classification system, without personal judgements or biases. This would determine the form of entrance to the tournaments and would mark an objective sample on the performance of the players. Twelve months after the creation of the ATP, Ilie Nastase became the first No. 1.

The tournaments were initially divided into categories –A, B, C, etc.–, which allowed the event organizers to select the players according to their ATP classification and determine the seeds.

The results of the tournaments were sent to the ATP where they were processed, then, the ranking was generated in huge perforated sheets and once a month they published the new ranking of players.

In the following years, after 11 publications in 1973, the ATP Ranking began to be published more frequently –1974 (11), '75 (13), '76 (23), '77 (34), 78 (40)– until 1979, when it was published once a week, 43 in the season.

The international ATP Rankings of August 23, 1973 was an average system, points earned over a period of 52 weeks were accumulated and then divided by the total number of tournaments played (with a minimum divisor of 12). The tournaments awarded points according to the prizes (minimum 25 thousand Dollars), the size of the table and its difficulty. The merit-based system was supported by the players.

Following several changes to the points counting system, in the year 2000 and in order to encourage greater participation in Grand Slams and in the series of nine most important tournaments, of the ATP (now known as the ATP World Tour Masters 1000), the qualifying or classification system began counting 18 events for most players. The 13 results of the Grand Slams and ATP World Tour Masters 1000 tournaments would count, as well as the top five performances of one player in the events of the International Series (now ATP World Tour 250 and 500).

## 2.2. ATP Ranking System

The Emirates ATP Rankings are the unbiased historical method based on merits to determine acceptance and seedling in all tournaments for singles and doubles.

As explained in [2, 3], the period that the Emirates ATP Rankings take is the last 52 weeks, however they do not take all the tournaments in which a player participates during that period. The ATP Ranking is based on the total points scored by one player in the four (4) Grand Slams, eight (8) of the nine (9) ATP Masters 1000, and their best six (6) results from the rest of the ATP tournaments in which he participated. For non-Top 30 players who do not directly qualify for the ATP Masters 1000 and Grand Slam tournaments, if they do not play one of them, a seventh tournament is taken into account for the rest of the tournaments played during the year. At the end of the season, the London-based ATP World Tour Final is played, where the top seven (7) players are automatically ranked as well as the winner of one of the Grand Slam tournaments (if he is already qualified, the eighth player as per the ATP Ranking also qualifies).

| Level/Round | W | F | SF | QF | 16 | 32 | 64 | 128 |
|---|---|---|---|---|---|---|---|---|
| Grand Slams | 2000 | 1200 | 750 | 360 | 180 | 90 | 45 | 10 |
| Masters 1000 | 1000 | 600 | 360 | 180 | 90 | 45 | 10[25] | [10] |
| ATP 500 | 500 | 300 | 180 | 90 | 45 | 20 | 0 | |
| ATP 250 | 250 | 150 | 90 | 45 | 20 | [10] | 0 | |

Scoring system per round achieved in each tournament type.

## 2.3. Critics to the ATP Ranking

According to [4], there are many critics of players to the scoring system of the ATP Ranking, many of them made by the most important players like Rafael Nadal.

The current ATP Ranking holds measurements over 52 weeks, which can make the ranking somewhat inconsistent week by week. For example, in the hypothetical case where a player wins an ATP Masters 1000, he would accumulate those 1000 points for 51 weeks from the end, but if he does not repeat the title, 53 weeks after receiving the trophy, he will no longer have those 1000 points. The potential oscillation between two positions in the ranking with such close dates does not really reflect the true ranking of the players.

On the other hand, all tournaments of the same level add the same amount of points, regardless of the players who participate in it. When counting only 6 of the tournaments not considered "big", a player can participate in as many tournaments as he can, aimed at increasing his score and obtaining the possibility of being pre-qualified in "big" tournaments and obtaining a more favorable classification.

Finally, the current ranking system penalizes those who get an unfortunate draw in big tournaments. If a player always has to play against a Top 10 in the first round, his chances of moving up his ranking are small.

### 2.4. ATP as a predictive model

If we look at the ATP Ranking not only as a qualifying ranking but also as a predictive ranking, we could try to predict who may be the winner before each match based on the players' rank.

Considering the player with the lowest number in the ranking as the player with the best present at that time, we evaluated in games from 2005 to 2013 how this predictive model worked.

Therefore, we obtain:

| YEAR | ATP |
|---|---|
| 2005 | 66.430% |
| 2006 | 66.494% |
| 2007 | 65.455% |
| 2008 | 66.990% |
| 2009 | 68.072% |
| 2010 | 66.641% |
| 2011 | 67.758% |
| 2012 | 67.908% |
| 2013 | 65.892% |
| **TOTAL AVG.** | **66.849%** |

Efficiency of the ATP Ranking as a predictive system

We can argue that the ATP Ranking matches in approximately 2/3 of the matches as established. It can also be interpreted that in about 2/3 of the cases the best positioned player is the winner of the match.

The way it is being evaluated is by counting the number of matches where the best placed player won, i.e. a HIT, over the total number of games played during the period under evaluation.

## 3. PAGERANK

### 3.1. Introduction

PageRank is an algorithm used by Google Search to rank websites in order to valuate and differentiate them according to their importance.

PageRank measures the "authority" of a page on several specific topics. The more authority, the more chances this website will have to appear in the top positions of related searches.

It is measured by knowing the number of links that point to that page, along with the authority of the page that links it and the way it does it.

Each link counts as a vote or a recommendation. And the recommendation is as important as the one who recommends it and how he does it. That is to say, that if the adviser has greater weight, his recommendation will be of greater importance as compared to the recommendation of another site.

### 3.2. PageRank History

Faced with the immense growth of the Internet and the large number of heterogeneous pages on the web, search engines were forced to improve their rankings to provide inexperienced users with the best answers.

In their beginnings, search engines functioned as websites crawlers in which the terms found on the pages were listed. When a word was searched in the search engine, it was searched in the term listings and pages where they appeared. The sorting method was based on the number of times a word was mentioned, which could lead to spam cases where everyone on their site regardless of topic repeated the most searched words to be able to bring traffic to their website.

Analyzing this and considering that although the number of sites had increased, they were still hypertexts that provided additional information, a ranking called PageRank was created as a method to compute each existing page as a node of a great graph that formed the web. Each node had a degree of importance that corresponded to whether it was an important page. This was determined based on how many links reached each node.

The algorithm was patented by Google founders Larry Page and Sergey Brin when they were doctoral students at Stanford University in 1999 [6, 8]. Page was a PhD student of Terry Winograd, who seems to have encouraged him to work on PageRank, and Brin was a Ph.D. student of Jeffrey D. Ullman, the famous author of the compiler book with Aho, although he soon joined Page to work on "more interesting" topics than his director offered.

This algorithm was used to power a new search engine called BackRub, which later became known as Google.

The Internet is today what it is thanks to the great effort they had to order the information in a relevant way in an environment in which the big portals had sold the results of the searches to the highest bidder.

Out of the 24 million web pages that its first version managed to index, the Internet has grown today to surpass, according to estimates, the 4,000 million different addresses. Since the algorithm has to find not only the first-order links that a page receives, but also those of higher orders, the real problem is not found in the original idea of how to measure relevance on the Internet, but in how to index the highest percentage of existing Internet sites and to evaluate the links that enter and leave each site.

## 3.3. Definition and Algorithm

Based on the concept of centrality of a graph, in which a value is given to a node based on its location, it is possible to determine which node is the most influential of the graph in its adjacency matrix. A node with a high value indicates that such node is connected to other nodes with considerable value as well; the higher the value of the node, the more to the center its location.

Using these concepts comes the creation of PageRank which, precisely, is a function that gives a value to each node corresponding to a web page.

The web is considered as a directed graph connected by the hyperlinks of the pages that are pointed, i.e., each edge represents a link between the outgoing node to the incoming node. Any random site is taken as a starting point to click on the different links, navigating this way from page to page.

The PageRank value of a page corresponds to the frequency with which any browser visits a page. The more time a user spends on a page, the more important the PageRank on that page.

Due to the current size of the web, Google search engine uses an approximate iterative value of PageRank. This means that each page is assigned an initial PageRank value, and then the PageRank of all pages is calculated with cyclic calculations based on the formula of the PageRank algorithm.

On a more formal note, as stated in [18], the general value of the PageRank of any page can be represented as:

Where $u$ is a website, $F_u$ is the set of pages pointed by $u$ and $B_u$ is the set of pages pointing at $u$. $N_u = |F_u|$ is the number of links of $u$ and $c$ the factor used for normalization.

We started by defining a simple *R* ranking as a simplified version of PageRank:

$$R(u) = c \sum_{v \in B_u} \frac{R(v)}{N_v}$$

The most popular pages are those that receive more edges from other pages or nodes of the graph. An important feature of the PageRank algorithm is that it assigns weights to the edges. These are taken into account as the main tool for the final calculation of the ranking.

In the algorithm of PageRank each site counts as a link, however the weights of the edges accumulate as the nodes of the graph are connected.

The weight of the page is calculated by adding the weights of the links it receives, while the weight of the edges has the value of the division of the weight of the page for each link to which it refers.

This causes the most important pages to have a greater influence on the value of the PageRank than the less popular pages.

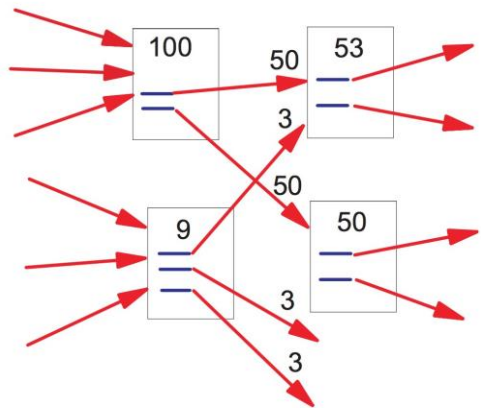

Calculation of PageRank with weight in the edges

Once the weight of each edge is calculated, an adjacency matrix is generated to calculate the positions that each page will take in the ranking.

## 3.4. PageRank Uses

The PageRank algorithm is very useful as a popularity indicator. Given any directed graph, it is able to indicate which node is the most important and thus be able to build a ranking on each of them.

As stated in [7], the PageRank algorithm is not only used as one of the main factors for Google site positioning. Considering its rankings ability, it is also used by large companies to build particular rankings, regardless of search engines.

For example:

- **Twitter tweets ranking**: It is done by relating the mentions from one user to another, thus forming a graph where the nodes are the users and each mention indicates an edge directed from one user to the other.
- **Recommendation systems**: It can be used in recommendation systems based on products consumed by a user. For example, Netflix can recommend movies based on cases of similar users.
- **Social networking friends suggestion**: Considering that each user feeds and is fed from other users, a PageRank system can be generated to suggest new friends based on those already connected.

## 3.5.  PageRank as a tennis ranking

Based on [9, 12, 13], there are different alternatives to the ATP Ranking in which the main objective is to get to a better reflection of who the best players are today, and in which position each one is located.

One of the algorithms that can be taken as a model to generate a new ranking is PageRank.

Based on the paper *On the (Page)Ranking of Professional Tennis Players*, written by Dingle, Knottenbelt and Spanias [11], a new tennis ranking is generated, where each node is a player and a directed edge of connection is traced between nodes given the result of a match. When one player wins another, an edge of weight 1 (one) is added from the losing node to the winning node.

Once all the games have been evaluated, the PageRank algorithm is applied on that graph and thus a ranking is obtained based exclusively on the history of the results of the matches, unlike the ATP Ranking where the tournament instance is evaluated.

We generate a graph based on the results of the matches of a tournament.

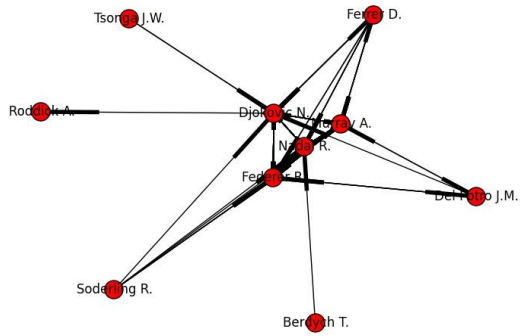

From this graph, the PageRank manages to generate a ranking for the participating players.

| Position | Player |
|---|---|
| 1 | Djokovic N. |
| 2 | Nadal R. |
| 3 | Federer, R. |
| 4 | Murray A. |
| 5 | Ferrer D. |
| 6 | Del Potro J. M. |
| 7 | Rodick A. |
| 8 | Tsonga J. W. |
| 9 | Soderling R. |
| 10 | Berdych T. |

Finally, this method, unlike the ATP Ranking, also allows us to build a ranking with players from different times.

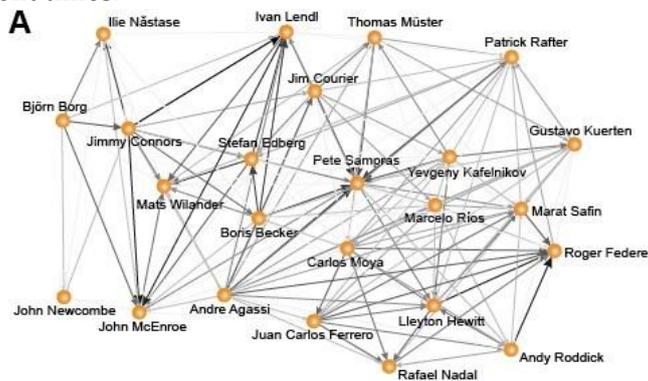

Graph generated by the results of ATP matches taken from [10]

## 3.6. PageRank v. ATP Ranking

Comparing the ATP Rankings versus the ranking generated through a PageRank algorithm we can observe that they can have many variations between the positioning assigned to some players by one system and by the other.

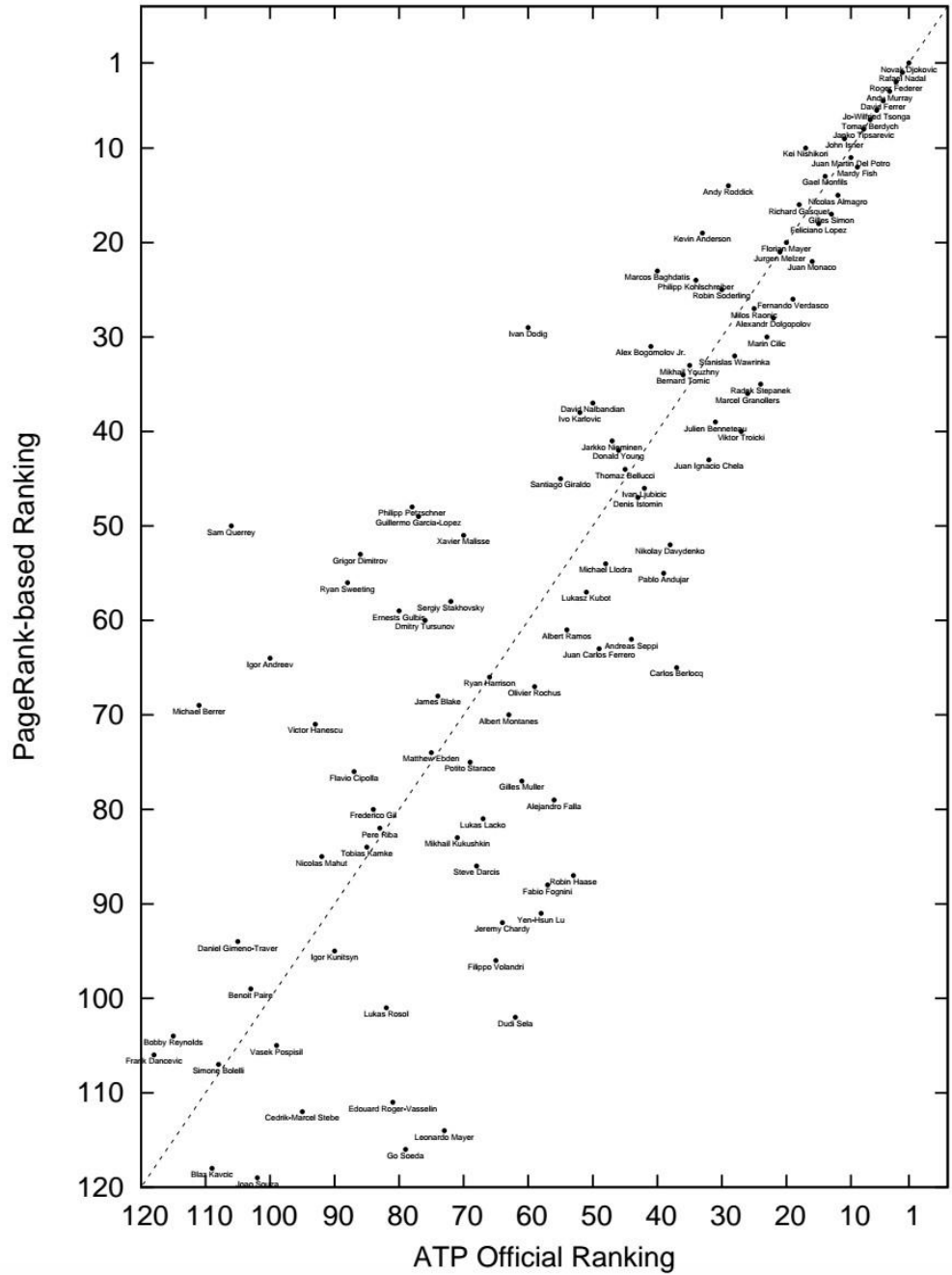

Comparative ranking ATP vs. PAGERANK extracted from the paper by Dingle, Knottenbelt and Spanias. [11]

| RANK | ATP | PAGERANK | DIFFERENCE |
|---|---|---|---|
| 1 | Djokovic | Djokovic | 0 |
| 2 | Nadal | Nadal | 0 |
| 3 | Murray | Federer | 4 |
| 4 | Ferrer | Murray | -1 |
| 5 | Berdych | Ferrer | -1 |
| 6 | Del Potro | Berdych | -1 |
| 7 | Federer | Tsonga | 1 |
| 8 | Tsonga | Del Potro | -2 |
| 9 | Gasquet | Isner | 8 |
| 10 | Wawrinka | Tipsarevic | 11 |
| 11 | Raonic | Almagro | 4 |
| 12 | Nishikori | Gasquet | -3 |
| 13 | Haas | Wawrinka | -3 |
| 14 | Janowicz | Raonic | -3 |
| 15 | Almagro | Simon | 1 |
| 16 | Simon | Nishikori | -4 |
| 26 | Lopez | Youzhny | -2 |
| 54 | Hanescu | Nalbandian | 50 |
| 55 | Andujar | Berlocq | -7 |
| 56 | Matosevic | Ramos | 19 |
| 57 | Delbonis | Giraldo | 28 |
| 58 | Stepanek | Hewitt | 8 |
| 75 | Ramos-Vinolas | Hanescu | -21 |
| 76 | Carreno Busta | Matosevic | -20 |
| 77 | Sela | Blake | 23 |

Comparison ATP Ranking vs. PAGERANK prior to US OPEN 2013

In the case of the chart we can see that when starting the US OPEN 2013 (Grand Slam Tournament) they have variations except in the first two places in the ranking.

For this tournament, once finished, the ATP rankings hit the results by 70.25% while the rankings put up by PageRank had a greater efficiency when achieving a positive 72.72% of the results.

Finally, we evaluated year-on-year performance of the PageRank in comparison to the ATP Ranking.

| AÑO | ATP | PAGERANK |
|---|---|---|
| 2005 | 66.430% | 66.785% |
| 2006 | 66.494% | 66.087% |
| 2007 | 65.455% | 66.939% |
| 2008 | 66.990% | 66.563% |
| 2009 | 68.072% | 69.516% |
| 2010 | 66.641% | 68.472% |
| 2011 | 67.758% | 67.524% |
| 2012 | 67.908% | 68.299% |
| 2013 | 65.892% | 68.010% |
| PROMEDIO TOTAL | 66.849% | 67.577% |

Comparison ATP Ranking vs. PAGERANK

As it can be seen, this new ranking system achieves greater precision if we evaluate it as a qualifying ranking.

That is to say, taking the results between the players as a reference and connecting them to be able to generate a ranking is much more effective at the time of predicting the result than evaluating only the instance of the tournament to which they arrive as does the ATP Ranking.

# 4. PARAMETRIC PAGERANK

## 4.1. Motivation

Given the good results that were obtained using the PageRank method above the ATP Ranking, we came up with the idea of looking for a better alternative that would improve both existing methods. This new method had to be able to counter the criticism of the players to the ranking used today and at the same time be more efficient and able to reflect the true position of tennis players in a ranking, as compared to the methods already evaluated.

As we have seen, the difference of the ATP Ranking in comparison with the PageRank ranking is that the first evaluates the players according to the instance they reached in the tournaments played, while the other evaluates the match-by-match results they have had in the last period.

That is why we ask ourselves: *Is it correct to evaluate only the instance of the tournament to which the players get? Is it correct to evaluate only 52 weeks back to know the present of a player? Do all matches won by a player over another have the same value? Is the context in which each victory is developed not relevant?*

Given these questions that the idea of combining both models arises. Considering PageRank as a more effective model than the current ATP Ranking, we use it as a base scheme to design a new ranking. In this new system, not only will the results between players be taken into account, but also other important attributes, such as the instance of the tournament in which the encounter took place (as in the ATP Ranking), the surface in which each match was played and the age of each one of those encounters.

We will look for the parameter that best represents each of these attributes and a weight will be assigned corresponding to each edge of the graph. Then the different parameters will be combined to achieve a ranking generated by the evaluation of all the abovementioned factors.

## 4.2. Data Set

As retrieved from www.tennis-data.co.uk, there is a set of data from each of the 923 tournaments played between 2000 and 2013.

By analyzing each of these tournaments, results can be taken from almost 40,000 matches with a detail of the type of surface of each game (hard, clay, grass), the date of each match, the importance of the tournament (ATP250, ATP500, MASTERS1000, Grand Slam, Master Cup).

The result of match set to set is also indicated, the original ATP Ranking of each player at the time of the game, and the amount of bets paid for the victory of each player.

All of this data was cured, tournament names, players and tournament type were unified and rolled into 3 tables of a MySql database for better reading. The tables are standardized by Players, Tournaments and Matches. (More info in the Appendix).

On the other hand, information was captured on the ATP website (www.atpworldtour.com) in which the top 100 ATP Rankings are processed since 1973.

## 4.3. Model

Similar to [11], a directed graph will be generated for each tournament corresponding to its matches.

Each edge will be strictly linked to the result of the match, and, by scrolling the list of matches, there will appear a number of assigned years, prior to the tournament to be evaluated, which in turn are added to a directed multigraph, from which we will finally calculate the PageRank.

Our proposal consists, unlike the aforementioned PageRank, of including a value as a weight of each edge, where each match will have a different value in our multigraph.

For such purpose, we will take into account 3 attributes that will help to calculate the corresponding weight for that match:

- Aging: It considers how old is the game under evaluation with respect to the tournament from which the ranking will be taken.
- Surface: It evaluates the difference of surface in the match under evaluation with respect to the tournament from which the ranking will be taken.
- Type of tournament and instance reached: It evaluates the type of tournament and the instance of that tournament in which the game was played just like ATP does.

As a development from this model, we will first analyze aging in years and tournaments that we will look back to be able then to evaluate each parameter individually and find the best value to generate a combination of them. In this way we will create a parametric PageRank more effective than the ATP ranking and the existing PageRank.

Finally, the weight of each bow will be indicated under the following formula [where PESO is weight, Arco is edge, Envejecimiento is aging, Superficie is surface and Instancia is instance]:

$$WEIGHT_{Edge} = WEIGHT_{Aging} * WEIGHT_{Surface} * WEIGHT_{Instance}$$

### 4.3.1.     Age

*Is it correct to take only 52 weeks as a reference to generate the ranking?*

As it was already mentioned, the ATP Ranking is generated based on the sum of points obtained in the tournaments that took place in the last 52 weeks.

In order to study if this was correct or if better results could be obtained if we looked at older matches, we evaluated different age parameters to know how many years backwards it is advisable to consider in order to obtain better results.

For this, each calculation of the different parameters that will be taken into account in the weight of the edge will be evaluated taking into account multiple years of age and choosing the one with better efficiency.

### 4.3.2.     Aging

*Is it correct to evaluate all matches in the history with the same weight?*

Unlike the ATP Ranking or the PageRank generated in the previous chapter, we will be evaluating many years back to be able to achieve a ranking more effective than the previous ones.
However, we believe it is not correct to weigh equally matches that were played long ago compared to more recent matches. That is why we will take the aging factor as one of our parameters. For such purpose, we will evaluate how a ranking will be calculated to give less weight to matches which are more distant in time and give a higher weight to those which have been played lately.

#### Exponential Decay

To represent the aging factor in the best way, we find in the Exponential Decay process a way of expressing the weight, in which the older matches had less weight than the new ones.

The Exponential Decay formula is represented as follows:
$$N(t) = N_0 e^{-\lambda t}.$$

Where *N(t)* is the weight to be assigned to the match represented in time *t*, which indicates the time difference between the tournament for which the ranking is being generated compared to the tournament under evaluation.

The Exponential Decay model is characterized by its rapid decrease, thus achieving very low values for those older matches.

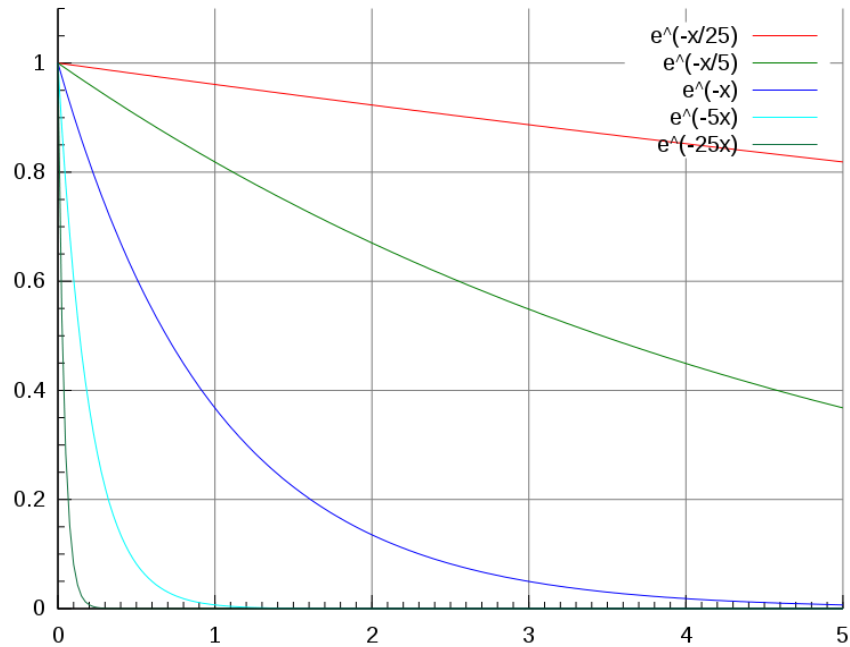

By modifying the corresponding $\lambda$ the drop speed can be adjusted.

Considering this, we evaluate the best value of $\lambda$ and evaluate it with the new ranking system generated.

For this we will take as age of the tournaments 3 and 5 years, that is to say, the graph will be composed by nodes and edges corresponding to matches as old as these.

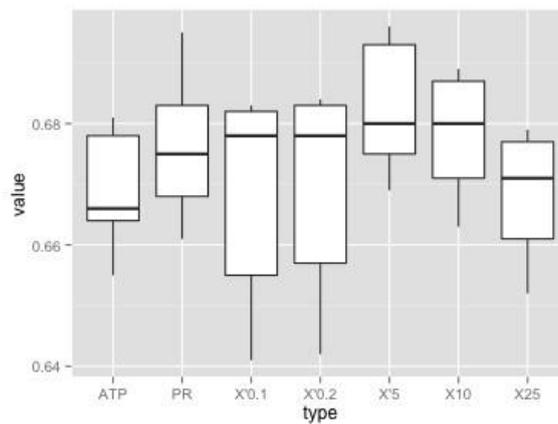

| AÑO | ATP | PAGERANK | ANTIGUEDAD DE 3 AÑOS | | | | |
|---|---|---|---|---|---|---|---|
| | | | 0.1 | 0.2 | 5 | 10 | 25 |
| PROMEDIO TOTAL | 66.849% | 67.577% | 66.820% | 66.913% | 68.237% | 67.824% | 66.836% |

Comparison of the ATP Ranking and generated PageRank, as compared to the new model taking 3 years of age.

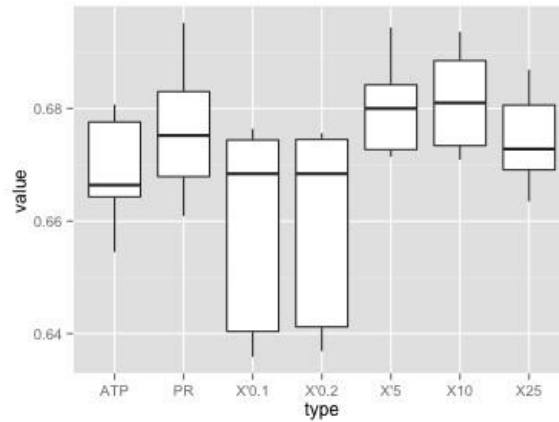

| AÑO | ATP | PAGERANK | ANTIGUEDAD DE 5 AÑOS | | | | |
|---|---|---|---|---|---|---|---|
| | | | 0.1 | 0.2 | 5 | 10 | 25 |
| PROMEDIO TOTAL | 66.849% | 67.577% | 65.849% | 65.899% | 68.008% | 68.150% | 67.443% |

Comparison of the ATP Ranking and generated PageRank, as compared to the new model taking 5 years of age.

As it can be seen, if we set an age of 3 years with a value of $\lambda$ in the Exponential Decay of -5, we obtain an improvement as a predictive ranking of 1.3% with respect to the ATP Ranking and of 0.8% with respect to the PageRank in which all its edges have equal weights.

### 4.3.3. Type of tournament and instance reached

*Is it the same to win a match of a first round than a final? And to play an ATP 250 than a Grand Slam?*

As we all know, tennis is a sport of high individual rivalry. Each match is loaded with pressure and the state of mind is different according to the importance of the match. That is why we consider it is not correct to assign the same weight to matches of different tournaments and matches in different rounds.

In the same way ATP does, we will try to generate the weight of the graph's edge according to the round of the tournament in which the match evaluated is taking place in order to generate the PageRank.

For that we will use a formula that has a quantity of points assigned similar to the one delivered by the ATP. The formula is:

$$Peso_{instancia} = \frac{2000/\lambda^{Valor_{instancia}-1}}{2000}$$

Where the value of the round reached in that tournament is the numeric value that represents the number of round reached depending on the type of tournament.

We will look then for the $\lambda$ between 1 and 2 that behaves better as a predictive ranking and at the same time comes close to the ATP ranking.

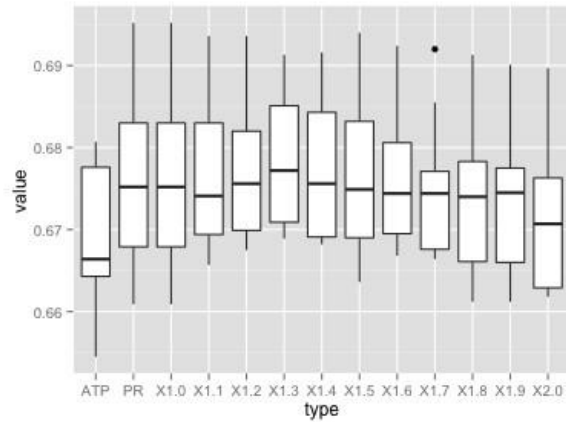

| AÑO            | ATP     | PAGERANK | 1       | 1.1     | 1.2     | 1.3     | 1.4     | 1.5     | 1.6     | 1.7     | 1.8     | 1.9     | 2       |
|----------------|---------|----------|---------|---------|---------|---------|---------|---------|---------|---------|---------|---------|---------|
| PROMEDIO TOTAL | 66.849% | 67.577%  | 67.578% | 67.639% | 67.716% | 67.763% | 67.687% | 67.573% | 67.536% | 67.482% | 67.369% | 67.341% | 67.249% |

Evaluation of the corresponding λ to score the round of the tournament reached.

It can be observed how the λ in 1.3 it is barely above than what the existent PageRank manages to predict although it has a much less variance. By comparison we did achieved a difference of at least 1 point in respect of the ATP ranking.

### 4.3.4.   Surface

*How much is the influence in the matches history to play in the same surfaces opposite to different surfaces?*

There are three types of surfaces in the ATP circuit: *Hard, Clay, Grass*. Each one of those surfaces has different characteristics.

In clay, for example, the ball has a much slower speed and it is easier to slide through the court. In cement, the ball goes at a higher speed and with more pronounced bounces, while in grass, besides having a great speed, the ball does not usually gain height.

Given these characteristics, there are players that know how to better develop their game in some surfaces, while in others they do not achieve a good performance. And because of this characteristic, it came up the idea of differentiating the weight of an edge that represents each match of the history, based on the surface of that game regarding the surface of the tournament we are trying to predict.

We generated a matrix in which the edges corresponding to matches played in the same surface than the tournament to be predicted the highest value that is 1 (one) is assigned as weight, while the edges corresponding to matches played in a surface different than the one of the tournament to be predicted are assigned a weight that we will calculate.

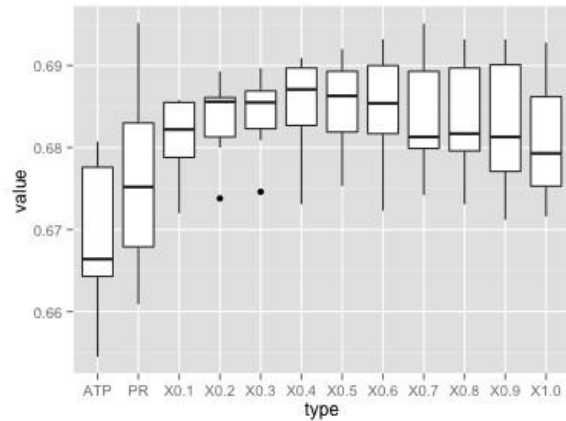

| AÑO | ATP | PAGERANK | 0.1 | 0.2 | 0.3 | 0.4 | 0.5 | 0.6 | 0.7 | 0.8 | 0.9 | 1 |
|---|---|---|---|---|---|---|---|---|---|---|---|---|
| PROMEDIO TOTAL | 66.849% | 67.577% | 68.141% | 68.394% | 68.424% | 68.520% | 68.557% | 68.508% | 68.393% | 68.347% | 68.242% | 68.105% |

Evaluation of surfaces under different parameter values in different surfaces

As it can be observed, by evaluating the rest of the surfaces with a parameter of 0.5 an efficiency of 1.7% greater with respect to the ATP ranking while we surpassed the static PageRank for almost 1%.

Knowing these results, we evaluated the results differentiating the surface. It is worth to note that most of the matches of the season take place in the hard court, followed by clay, while only a few of the weeks of the season are played in grass.

We can observe that even though each surface behaves better before a different parameter, in each one of them with the best choice our model behaves better than the ATP ranking in any of the surfaces.

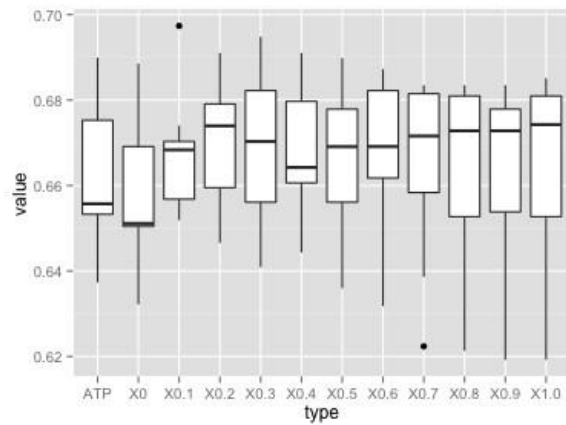

| | CLAY | | | | | | | | | | | |
|---|---|---|---|---|---|---|---|---|---|---|---|---|
| | ATP | 0 | 0.1 | 0.2 | 0.3 | 0.4 | 0.5 | 0.6 | 0.7 | 0.8 | 0.9 | 1 |
| PROMEDIO TOTAL | 66.083% | 65.698% | 66.709% | 66.925% | 66.867% | 66.711% | 66.557% | 66.650% | 66.432% | 66.265% | 66.180% | 66.273% |

Evaluation of clay surfaces under different parameter values in different surfaces

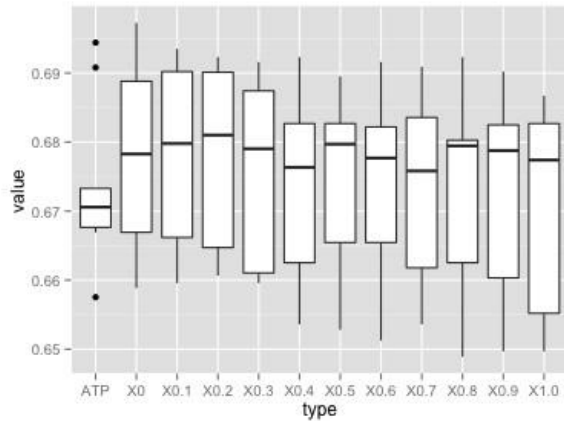

| | HARD | | | | | | | | | | |
|---|---|---|---|---|---|---|---|---|---|---|---|
| | ATP | 0 | 0.1 | 0.2 | 0.3 | 0.4 | 0.5 | 0.6 | 0.7 | 0.8 | 0.9 | 1 |
| PROMEDIO TOTAL | 67.353% | 67.861% | 67.869% | 67.814% | 67.657% | 67.430% | 67.439% | 67.298% | 67.239% | 67.246% | 67.150% | 67.053% |

Evaluation of hard surfaces under different parameter values in different surfaces

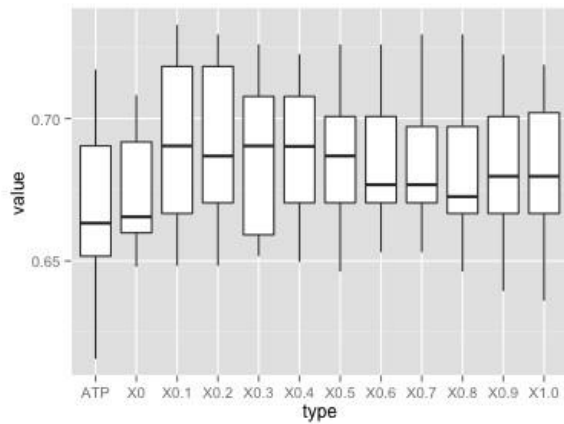

| | GRASS | | | | | | | | | | |
|---|---|---|---|---|---|---|---|---|---|---|---|
| | ATP | 0 | 0.1 | 0.2 | 0.3 | 0.4 | 0.5 | 0.6 | 0.7 | 0.8 | 0.9 | 1 |
| PROMEDIO TOTAL | 67.0181% | 67.4533% | 69.3553% | 68.9427% | 68.8134% | 68.9709% | 68.5859% | 68.3960% | 68.3967% | 68.2814% | 68.2428% | 68.1278% |

Evaluation of grass surfaces under different parameter values in different surfaces

### 4.3.5.     Parameter combination

With the certainty that our model behaves better than the previously mentioned, we look for the best combination of parameters in order to obtain a more efficient result than when we observe the attributes in particular.

For that we will combine the parameters and we will use the multiplication of them like weight of the edge within the formed directed multigraph from which the PageRank will be calculated.

As a first intuition we tried the combination of the best parameters that we got particularly. Meaning that we took a 5-year aging with an Exponential Decay of -5 as λ, we will use 1.3, and as surface 0.5 to calculate the tournament round reached.

As a result of this combination we have:

| AÑO | ATP | PAGERANK | PAGERANK PARAMÉTRICO |
|---|---|---|---|
| 2005 | 66.430% | 66.785% | 68.632% |
| 2006 | 66.494% | 66.087% | 67.530% |
| 2007 | 65.455% | 66.939% | 68.163% |
| 2008 | 66.990% | 66.563% | 68.191% |
| 2009 | 68.072% | 69.516% | 69.204% |
| 2010 | 66.641% | 68.472% | 68.784% |
| 2011 | 67.758% | 67.524% | 69.162% |
| 2012 | 67.908% | 68.299% | 68.926% |
| 2013 | 65.892% | 68.010% | 68.419% |
| PROMEDIO TOTAL | 66.849% | 67.577% | 68.557% |

Evaluation with the best parameter calculated individually

In spite of obtaining a better result, we chose to look in every parameter the best combination possible to obtain an even better result.

For that we put together an algorithm that will make a greedy search in which within a vector three of the four values will be fixated that we are repeating and the other we will repeat it until we find the best result, then we will go changing the vector fixing different values. We will do this until after several runs, we will manage to obtain the best result, meaning that after trying with all the parameters again, none of them were modified, for which our vector is a local optimum for the problem.

We evaluated the best parameters by generating a testing set composed of the matches of 90 tournaments, 10 by year, and as a result, the combination:

- Age: 4 years

- Aging: -5 as value of Exponential Decay

- Surface: 0.3 as value for different surfaces

- Tournament and round: 1.7 as value of λ to assign a score to the played round of the tournament

Combining these values we obtained:

| AÑO | ATP | PAGERANK | PAGERANK PARAMÉTRICO |
|---|---|---|---|
| 2005 | 66.430% | 66.785% | 70.755% |
| 2006 | 66.494% | 66.087% | 69.143% |
| 2007 | 65.455% | 66.939% | 69.767% |
| 2008 | 66.990% | 66.563% | 67.290% |
| 2009 | 68.072% | 69.516% | 70.190% |
| 2010 | 66.641% | 68.472% | 71.464% |
| 2011 | 67.758% | 67.524% | 69.231% |
| 2012 | 67.908% | 68.299% | 69.799% |
| 2013 | 65.892% | 68.010% | 70.915% |
| **PROMEDIO TOTAL** | **66.849%** | **67.577%** | **69.839%** |

Evaluation with the best parameter calculated after obtaining the best results evaluated with the algorithm

As it can be seen, an approximation to 70% of hits in the evaluated tournaments, achieving 3 points of improvement with respect of the hits of the ATP in its weekly ranking and 2.2 with respect to the existent PageRank model.

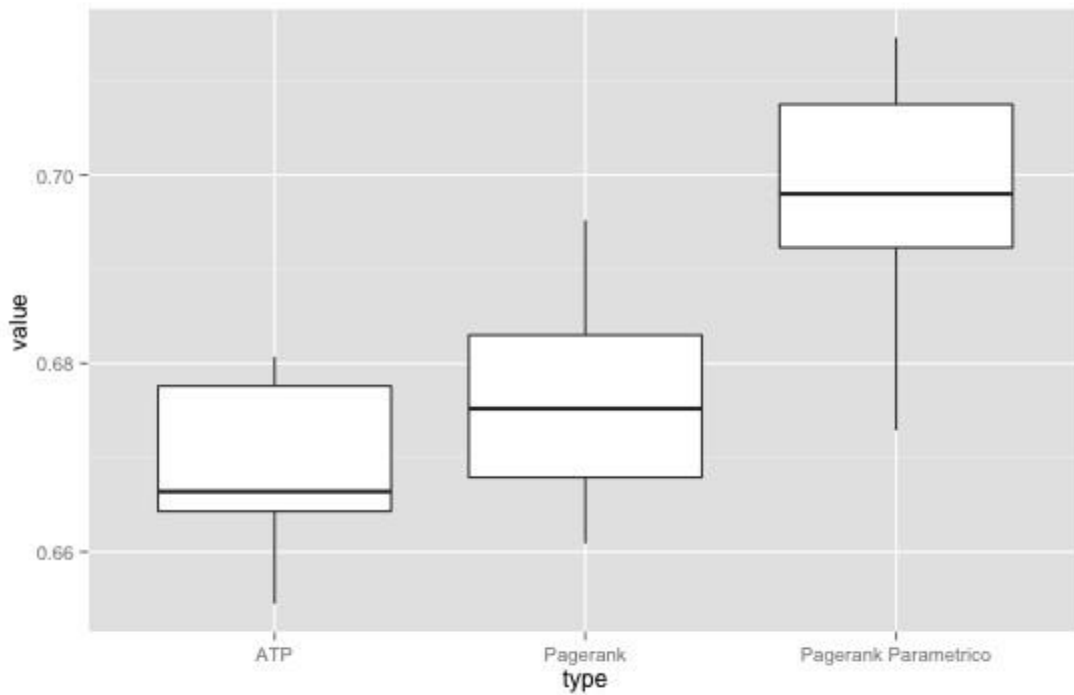

Boxplot of predictive rankings compared, ATP, PageRank and Parametric PageRank

In the *boxplot* type graph we can notice that regarding the median of the Parametric Pagerank is above of the rest of the rankings studied. It can be noticed that the minimum result obtained with the suggested model, reaches the mean of the Pagerank existent and surpasses the ATP ranking.

It can also be observed that the suggested model as the table shows, has around 70% of hits, counting with results that even surpass this calculation.

By using the statistic test ANOVA (Analysis of Variance) we could confirm that there is not a considerable difference in the effectiveness shown comparing the ATP with the existent Pagerank (p = 0.14). Meanwhile we see a great difference between the results obtained by the ATP ranking and the Parametric Pagerank (p = 0.000024) and also an important difference regarding the existent Pagerank
(p = 0.00079)

### 4.3.6. Results of itemized parameters

Before the finding that the presented model achieved a predicting character better than the rest of the existent models, we started to investigate if it behaved in a similar way in different laid out existent situations in the tennis tournaments and matches frame.

That is why we itemized the matches based on the surface, ranking of players and type of tournament in order to find in which frames our model had better predictions and in which had worse.

- Surface: In these cases we trained with all the surfaces but only tested on one particular surface

| HARD | | | |
|---|---|---|---|
| AÑO | ATP | Pagerank | Pagerank Paramétrico |
| 2005 | 66.764% | 66.178% | 68.452% |
| 2006 | 67.057% | 65.886% | 67.716% |
| 2007 | 66.768% | 67.525% | 69.114% |
| 2008 | 65.752% | 65.752% | 66.850% |
| 2009 | 69.441% | 69.930% | 70.350% |
| 2010 | 67.296% | 68.577% | 68.847% |
| 2011 | 67.329% | 66.918% | 69.247% |
| 2012 | 69.080% | 69.012% | 69.621% |
| 2013 | 66.691% | 68.769% | 68.398% |
| Promedio general | 67.353% | 67.616% | 68.733% |

Comparison by hard court

| CLAY | | | |
|---|---|---|---|
| AÑO | ATP | Pagerank | Pagerank Paramétrico |
| 2005 | 64.749% | 66.736% | 67.678% |
| 2006 | 65.568% | 64.659% | 65.682% |
| 2007 | 64.253% | 67.647% | 67.647% |
| 2008 | 68.990% | 67.548% | 68.510% |
| 2009 | 65.806% | 69.250% | 68.635% |
| 2010 | 65.574% | 68.096% | 70.366% |
| 2011 | 67.528% | 68.020% | 68.881% |
| 2012 | 67.970% | 67.970% | 68.222% |
| 2013 | 65.328% | 67.153% | 67.762% |
| Promedio general | 66.196% | 67.453% | 68.153% |

Comparison by clay

| GRASS | | | |
|---|---|---|---|
| **AÑO** | **ATP** | **Pagerank** | **Pagerank Paramétrico** |
| 2005 | 71.717% | 71.044% | 72.391% |
| 2006 | 65.169% | 70.037% | 69.288% |
| 2007 | 66.327% | 64.626% | 65.986% |
| 2008 | 69.039% | 70.819% | 71.886% |
| 2009 | 68.858% | 70.242% | 70.588% |
| 2010 | 66.207% | 68.966% | 66.897% |
| 2011 | 70.548% | 69.178% | 69.521% |
| 2012 | 61.566% | 65.480% | 67.616% |
| 2013 | 63.732% | 66.901% | 67.254% |
| **Promedio general** | 67.018% | 68.588% | 69.047% |

Comparison by grass

As it can be observed, our model achieves better results in every surface, achieving a greater possibility of prediction in those matches played in grass where it comes closer to the general average of the model and achieving 2% more than in the ATP ranking.

It can be also observed that both on hard court and in clay, our model is the one with the most hits even though the difference with other models is smaller.

- Ranking: In these cases we will evaluate the matches between similar ranking players and that only belong to that delimited ranking group.

| 1 a 10 | | | |
|---|---|---|---|
| **AÑO** | **ATP** | **Pagerank** | **Pagerank Parametrico** |
| 2005 | 70.270% | 78.049% | 76.60% |
| 2006 | 68.085% | 71.698% | 71.11% |
| 2007 | 66.667% | 69.565% | 68.42% |
| 2008 | 60.345% | 64.815% | 64.91% |
| 2009 | 63.529% | 66.250% | 59.15% |
| 2010 | 63.462% | 62.963% | 68.25% |
| 2011 | 67.568% | 65.217% | 69.62% |
| 2012 | 76.471% | 73.913% | 78.16% |
| 2013 | 58.974% | 71.642% | 67.11% |
| **Promedio general** | 66.152% | 69.346% | 69.260% |

Comparison matches between top ten

| 11 a 50 | | | |
|---|---|---|---|
| **AÑO** | **ATP** | **Pagerank** | **Pagerank Parametrico** |
| 2005 | 60.172% | 55.655% | 58.69% |
| 2006 | 61.165% | 53.041% | 58.36% |
| 2007 | 57.547% | 59.281% | 60.57% |
| 2008 | 55.963% | 50.485% | 56.13% |
| 2009 | 62.162% | 64.151% | 60.47% |
| 2010 | 55.799% | 57.049% | 57.05% |
| 2011 | 55.848% | 57.317% | 56.16% |
| 2012 | 58.333% | 55.128% | 58.28% |
| 2013 | 53.333% | 55.455% | 58.07% |
| **Promedio general** | 57.814% | 56.396% | 58.200% |

Comparison matches of the ranking 11 to 50

| 50 en Adelante | | | |
|---|---|---|---|
| AÑO | ATP | Pagerank | Pagerank Parametrico |
| 2005 | 58.100% | 59.538% | 59.848% |
| 2006 | 59.091% | 58.142% | 59.783% |
| 2007 | 56.364% | 60.358% | 58.978% |
| 2008 | 58.534% | 59.463% | 58.518% |
| 2009 | 56.377% | 56.364% | 58.177% |
| 2010 | 56.855% | 60.156% | 60.614% |
| 2011 | 60.931% | 59.814% | 61.485% |
| 2012 | 56.831% | 58.413% | 57.517% |
| 2013 | 56.825% | 60.699% | 58.742% |
| Promedio general | 57.767% | 59.216% | 59.296% |

Comparison matches of the ranking 50 and on

In this comparison we can observe that in the three cases our model surpasses the ATP ranking while it has an amount of hits similar with respect to the original Pagerank model.

In case of the matches between top ten players we achieved a difference greater than 3 points with respect of the ATP ranking. Which indicates that for matches between the great players, where the difference is much lower since the players involved are from the world top ten, our model accomplishes, as in the original Pagerank, to behave in a manner similar than what it does the ATP ranking.

For the rest of the cases, the model created achieves an improvement for two of the existent models, even though the difference is not substantial, our created ranking manages to achieve with a greater certainty who will be the winner of a match or a tournament if players in those positions play.

- Type of tournament: In this case we will evaluate the results according to the importance of the tournament that is being played. It is worth to note that the most important tournaments have a greater number of participants and a greater number of matches to be analyzed.

| ATP 250 | | | |
|---|---|---|---|
| AÑO | ATP | Pagerank | Pagerank Parametrico |
| 2005 | 67.180% | 67.180% | 67.488% |
| 2006 | 65.468% | 65.217% | 66.304% |
| 2007 | 63.365% | 66.509% | 66.116% |
| 2008 | 67.475% | 66.639% | 67.726% |
| 2009 | 64.677% | 65.576% | 66.067% |
| 2010 | 64.446% | 66.696% | 67.215% |
| 2011 | 66.725% | 66.115% | 67.509% |
| 2012 | 62.839% | 63.562% | 64.467% |
| 2013 | 62.285% | 64.642% | 64.823% |
| Promedio general | 64.940% | 65.793% | 66.413% |

Comparison tournaments ATP 250

| ATP 500 | | | |
|---|---|---|---|
| AÑO | ATP | Pagerank | Pagerank Parametrico |
| 2005 | 64.747% | 65.060% | 65.899% |
| 2006 | 62.264% | 66.667% | 64.623% |
| 2007 | 66.667% | 63.904% | 71.094% |
| 2008 | 63.037% | 64.419% | 62.751% |
| 2009 | 70.029% | 72.245% | 72.622% |
| 2010 | 65.775% | 67.286% | 67.914% |
| 2011 | 68.519% | 64.880% | 71.693% |
| 2012 | 70.516% | 69.944% | 73.464% |
| 2013 | 63.057% | 69.131% | 68.153% |
| Promedio general | 66.068% | 67.060% | 68.690% |

Comparison tournaments ATP 500

| MASTERS 1000 | | | |
|---|---|---|---|
| AÑO | ATP | Pagerank | Pagerank Parametrico |
| 2005 | 62.478% | 64.977% | 66.954% |
| 2006 | 67.185% | 61.792% | 68.566% |
| 2007 | 62.063% | 67.448% | 66.298% |
| 2008 | 63.670% | 61.318% | 64.607% |
| 2009 | 68.980% | 70.317% | 71.429% |
| 2010 | 63.941% | 66.578% | 68.401% |
| 2011 | 63.216% | 69.841% | 65.250% |
| 2012 | 69.388% | 72.482% | 70.315% |
| 2013 | 66.543% | 65.605% | 68.207% |
| Promedio general | 65.274% | 66.706% | 67.781% |

Comparison tournaments Masters 1000

| GRAND SLAM | | | |
|---|---|---|---|
| AÑO | ATP | Pagerank | Pagerank Parametrico |
| 2005 | 71.047% | 69.815% | 73.511% |
| 2006 | 71.837% | 70.816% | 70.612% |
| 2007 | 72.765% | 71.518% | 73.597% |
| 2008 | 72.279% | 72.279% | 74.127% |
| 2009 | 74.743% | 76.591% | 74.949% |
| 2010 | 74.948% | 74.741% | 74.741% |
| 2011 | 74.948% | 72.464% | 75.362% |
| 2012 | 74.845% | 73.196% | 73.402% |
| 2013 | 75.260% | 75.676% | 75.052% |
| Promedio general | 73.630% | 73.011% | 73.928% |

Comparison tournaments Grand Slam

| MASTERS CUP | | | |
|---|---|---|---|
| AÑO | ATP | Pagerank | Pagerank Parametrico |
| 2005 | 53.333% | 53.333% | 73.333% |
| 2006 | 66.667% | 80.000% | 73.333% |
| 2007 | 100.000% | 53.333% | 53.333% |
| 2008 | 66.667% | 73.333% | 73.333% |
| 2009 | 53.333% | 53.333% | 60.000% |
| 2010 | 86.667% | 93.333% | 86.667% |
| 2011 | 60.000% | 53.333% | 73.333% |
| 2012 | 93.333% | 86.667% | 86.667% |
| 2013 | 66.667% | 80.000% | 73.333% |
| Promedio general | 71.852% | 69.630% | 72.593% |

Comparison tournaments Masters Cup

As it can be observed, it is important to highlight that in the five types of tournament analyzed, our model accomplishes a better prediction than the rest of the analyzed models.

Excellent results can be observed of hits in the Grand Slam and Masters Cup tournaments, reaching almost 74% within the first ones and achieving a 72,6% in the second ones. However, it is not in these tournaments where a greater difference is achieved in terms of the existent rankings.

In the ATP 500 and Masters 1000 tournaments a difference is achieved with respect of the ATP of approximately 2.5%, what makes our model much more reliable.

# 5. ESTIMATION OF THE PROBABILITY OF VICTORY

## 5.1. What is the probability of victory?

As we have seen, we created a model capable of predicting the results of a given tournament with more effectiveness than what does today the ATP ranking, and the existent Pagerank model.

Before this scenario, the idea comes up of not only to be able to predict the result of a match based on the ranking, but also to determine with which odds the result of the player that is ranked better will be victorious in front of one ranked worse.

It is here where the odds of victory come up, that determines, given $r_1$ and $r_2$ (the rankings of the players that will play) and considering its difference, how firm are the odds that the better ranked triumphs.

## 5.2. How do we calculate it?

In order to find this chance of victory, we will first determine the efficiency that we had in our ranking if we grouped by the ranking difference. Meaning that for every game, we analyze not only the result, but also considering the ranking difference delivered by our model, how it behaved for that delta.

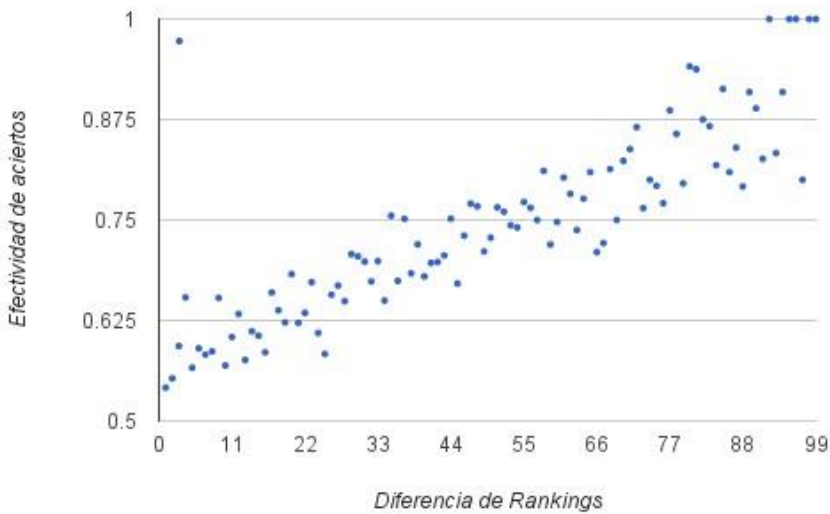

Evaluation of efficiency of victories based on ranking difference provided by our model

By analyzing the graphic obtained, we want to find the function that comes closer in the best possible way to every point of the graphic, for that we will use the regression model that will allow us to reach the objective in the best way.

Based on the paper [15] in which a similar scheme is used for soccer teams and considering that our graphic is similar to an exponential curve, we will look for the exponential that covers all the indicated points.

We will first define a model function that we consider more appropriate in base to the obtained. For that we will use a logistic that is characterized for representing the growth of organs from a small initial state, during which the growth is proportional to the size until the last stage where the size is similar to an asymptote.

Considering that our graphic starts in the initial point where the ranking difference between the players is minimal and then starts to grow up to the point where as the ranking difference grows the number of hits is considered permanent we can indicate that it has a behavior similar to an asymptote. That is why we can use the logistic function suggested in our case.

The function is of the type:

$$P_{Victory} = \frac{1}{1 + e^{\frac{r1-r2}{a}}}$$

Then we must take the following steps:

1. Estimate the parameters of the regression model. This process is called adjustment of the model to the data.

2. Test how good is the adjusted model. The result of this test may indicate if the model is reasonable or if the original adjustment must be modified.

By taking these steps, we can determine through the regression model that *a* = 45,321. In base of this parameter we get

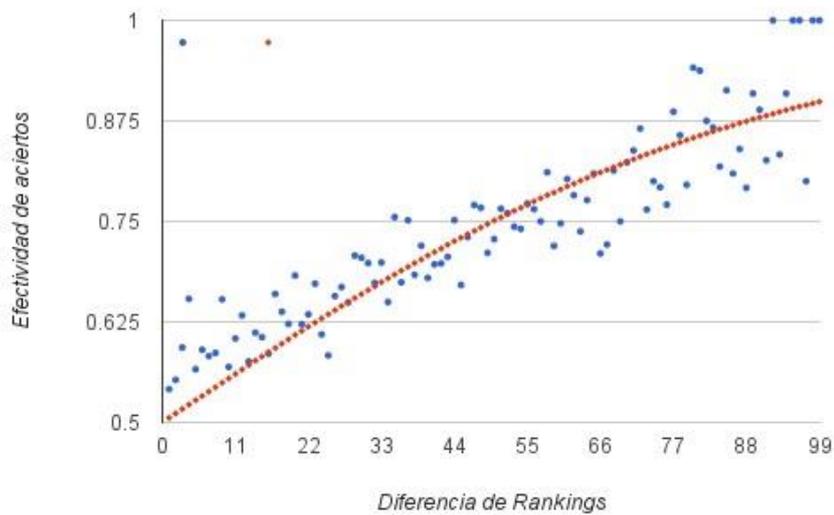

Evaluation of efficiency of victories based on ranking difference and regression curve

We can indicate that being this a reasonable curve we may continue to evaluate how it behaves based in our model.

## 5.3. Comparative of P(Victory)

Already knowing the function that represents our probability of victory, and reaffirming that the curve obtained manages a coherent form in front of the scored points, we focus on comparing how optimistic is our chance of victory in front of the probability of victory that may be obtained of the curve based in the ATP ranking or in the existent Pagerank model.

For that, we calculated for the top 100 of the players of each model which is the probability of victory of each difference.

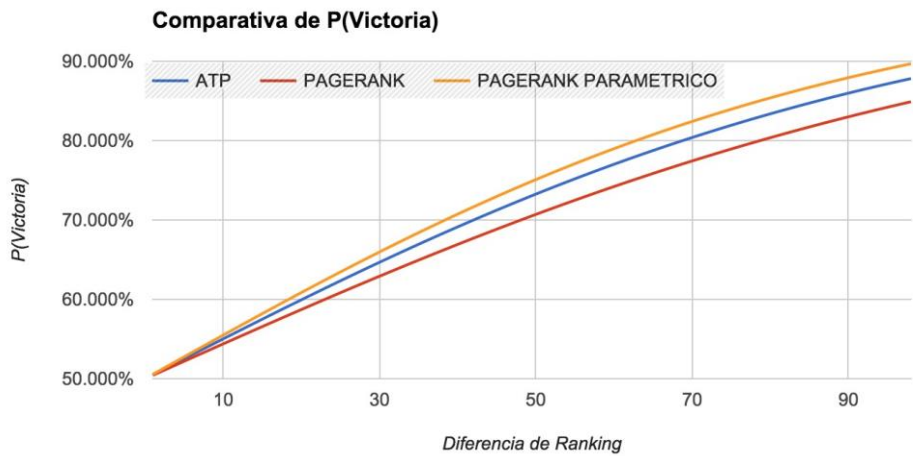

Comparison of victory probability of the existent models

It can be observed that the suggested model achieves a more optimistic character regarding the victory probability than the existent PageRank and much more even than the ATP ranking.

As the graphic shows, from the 20 positions of difference of ranking our suggested model already starts to differentiate itself from the ATP ranking, achieving a victory probability greater than 60% for that with a better ranking.

### 5.3.1. Specification based on specified parameters

As we have seen, in the general framework our generated ranking achieves a probability of victory more optimistic than the rest of the rankings. We can then look for the probability taking into consideration the attributed that were evaluated at the moment of generating the ranking.

Based on this a decision tree can be made, which edges can answer to the questions: In which surface it was played? What kind of tournament is the one being evaluated?.

By using the same comparison used in the generalized probability, we get to the pages of the combination *Surface - Type of tournament,* where our suggested ranking achieves to be more optimistic than the rest of the rankings.

Putting as an example some of the graphics of the pages obtained, it can be observed that unlike the generalized ranking, the optimism and the probability of victory of the different rankings manage to be more or less pronounced depending on the parameters with which they are calculated.

Then, given a match in which we know the positioning of each player in every one of the evaluated rankings, we can determine the odds that the players have of winning in that match by knowing also the type of tournament and the surface of that particular match.

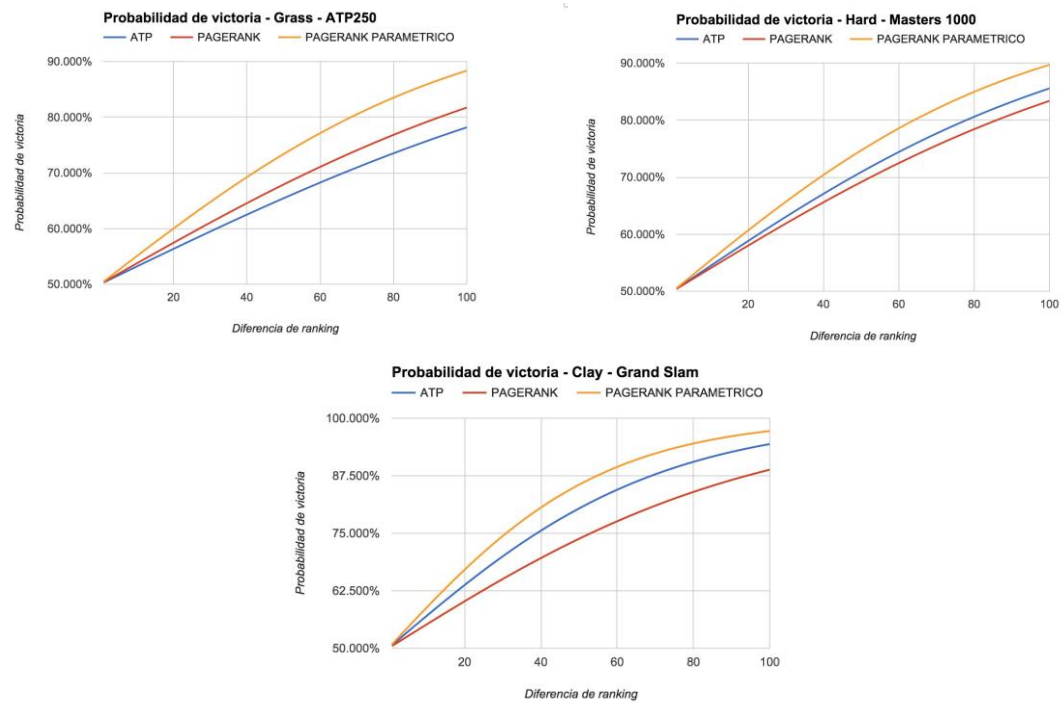

With that in mind, taking into account the numbers seen, if the ranking of the suggested model is taken as a reference, with a lesser difference of the ranking we can obtain a more important bet in which we can determine who will be the winner of the match, since it has proven to be more optimistic than the two other rankings seen.

## 5.4. Evaluation

Since we obtained the victory probabilities and since our model is more optimistic than the existent models, we focused on evaluating the performance on each one of them. For that we used the method of AUROC (Area under ROC).

As it is indicated in [16, 17] the analysis of ROC curves is a statistic method to determine the diagnostic accuracy of the tests performed, being used with three specific purposes:

- To determine the cut-off point of a continue scale in which the highest sensitivity and specificity is reached.

- To evaluate the discriminative capacity of the diagnose test, meaning, its capacity of differentiating with which odds a player will win a match with a given ranking difference.

- To compare the discriminative capacity of the diagnostic tests that express results in continuous scales.

The area under the ROC curve is an excellent performance global indicator of a diagnostic test since it makes possible to express it in a simple number.

Evaluating *P(victory)* for the 3 models we get:

| ATP | PAGERANK | PAGERANK PARAMETRICO |
|---|---|---|
| 0.7040 | 0.6776 | 0.7137 |

AUROC comparison

As it can be observed, our model is the one with best performance, however, unlike the efficiency shown in the previous chapter, the difference with the ATP ranking is smaller, while the difference with the Pagerank is considerable.

If we calculate the AUROC making each game to go in their corresponding edge in the decision tree, we found an improvement, even though it is not too big.

| ATP | PAGERANK | PAGERANK PARAMETRICO |
|---|---|---|
| 0.7070 | 0.6812 | 0.7168 |

AUROC comparison

The usage of the *P(victory)* allow us to widen the usage spectrum of our ranking. In this way it is possible to fix a satisfaction threshold according to our expectations of usage. In the "Future Work" we will expand this aspect.

# 6. FUTURE WORK

As we have seen, the rankings surround us in each of our everyday activities. With this work we have accomplished to introduce a new raking system in a sport environment like tennis, however, the fact that we created a new system and took the chance also to try to predict tennis results opens the door to many works to be made in the future.

- Application of other known rankings to the tennis environment.

    As it can be read in the book Who's the #1 [5], through history several ranking generating models had been developed [14], they are used today to classify teams in several sports. These methods, such as the Massey, the Keener, the Elo, among others, are mostly used to generate NFL or chess statistics, but not for tennis. And even though in our model different techniques were used that included their methods, since PageRank already has them, we believe that these methods can be applied purely to achieve different tennis rankings in order to discover which ranking best classifies the players today.

- Application of our parametric PageRank method in different sports.

    The generated method proved to behave better than the ATP ranking with which tennis players are classified nowadays. With these favorable results, a door opens to apply this ranking system in other sports where the results are zero sum and allow us to generate a PageRank in which the edges can be directed with different weights evaluating the characteristics of the played match.

    That is why the ranking can be applied in any racket sport and also in particular leagues, such as NFL or NBA.

    Regarding sports such as soccer, it can be applied as long as it is used for a national league, not international. The way in which this ranking is created, does not allow to know in these sports which one will be the best team in the world, since the matches between interleague teams are very few to contemplate that result.

- Application of our parametric PageRank method in different environments.

    Outside of the sport environment, it can also be applied in different environments such as fashion or culinary, or in any other field in which tournaments or contests are held throughout the year in which the participants can be ranked.

    For example, in the culinary contests that take place, a graph is generated exactly like if it was a tennis tournament, and then throughout the year the tournaments are gathered in order to indicate who is the best chef, applying the weight of the topic of the contest

held in the edges. The same application can be made with designers or even with videogame players.

- Tool for help in bets.

    With the results obtained, a great question comes up, can the generated ranking be used to gamble?

    Even though at first sight it appears that the numbers obtained are good and even with the P(victory) we can indicate in a given match which are the odds of a player or the other winning, this thesis opens the door to a more concrete analysis on the subject.

    We believe that the creation of a gambling assistant that can indicate which bets are the most reliable to make and which ones will deliver more money. Having the probability of victory in each game, we can indicate the minimum value of probability for which is advisable to make a bet.

    We can also try to indicate in the long term, the winners of the tournaments so that we can get a greater winning in a bet in case we win it.

    Other functions of this tool may be the simulation of how much money we can get by putting an initial amount and always betting to the winner that our ranking indicates based on the classification. We can make this by using our database since we have the bet values in the match history.

- Season planification for players.

    Using the victory probability and the generated model, we can work on the planning of the season of a player, finding what tournaments are best to play and arrive better prepared, prioritizing the ones that will potentially provide an improvement on the ranking.

    The same can be done by match, even by going to play knowing what the concrete chances of victory are.

    Seizing also the fact that our ranking can be easily differentiated by surface, type of tournament, and round, the strengths and weaknesses of the players can be known and see how they behave in a match with these attributes, which allows the trainers to customize the training routine in order to grow in those aspects and achieve substantial improvements in the performance of their players.

- Streaks

    A factor that can be added to the model we use to generate the ranking is the consideration of players' streaks. It happens in all seasons that there are moments of the year where the players that start a series of victories o defeats in a row that will make them either to escalate or to fall considerably in the ranking. This is something that does not behave well in the long run since many times they are isolated moments in the players' careers and do not reflect their level truly.

    Therefore, if the strikes are evaluated when generating the ranking, the matches in which these players were either victorious or defeated due to a strike can be given a different weight that allow our model not to be confused and to deliver a ranking that truly reflects in the classifying level, who is the best player for that moment, in a way that the results of the matches reflect the given classification.

# 7. CONCLUSIONS

Facing the idea of achieving a social ranking understood by all the tennis followers, the ATP ranking is exposed to constant complains of players and at the same time exposes new players to be benefitted with a good tournament in order to begin to progress in their careers.

At the same time, analyzing the results obtained, it can be observed that the ATP ranking is not capable enough to manage to predict with certainty who will be the winner of a match in case we based only in the positions.

With the objective of overcoming these obstacles, the idea comes up to create a new ranking that indicates which are the real victory odds of a player in the beginning of a new tournament. We tried to generate a new ranking that stands out because of the usage of the tournament characteristics for its generation.

At the beginning we implemented a ranking that uses Google's ordering algorithm, called PageRank, in order to achieve a player's ordering of positions in which the matches played by the players facing each other are more important than the round of a tournament reached by a player.

This type of ranking also, by the characteristic of the algorithm used, allows us to give more points to those players that won against the best players of the tournament and the circuit.

However, when comparing the prediction capability, it was only a little better than the current one, but not sufficient as to affirm that it is a best predictor.

It was then that we came up with the idea of implementing a ranking similar to the last one but where not only we will look at the result of the matches but we could also evaluate other attributes that these showed. Being one of those that took into account the ATP ranking. We thus created the Parametric Pagerank model, where we made a combination of the rankings already existent and also, we looked at other factors to evaluate it.

This new model behaves like the PageRank algorithm, but we add to each edge a different weight, based on the surface, round of the tournament, type of tournament, and age of the matches history evaluated.

As it can be observed, we have obtained great improvements in comparison with the existent models. Our model predicts the results of the matches and tournaments with respect to the ATP with roughly a 3% of improvement.

Our model also achieved a better capacity to predict any itemization we had made, by either matches between similar ranked players, type of tournament, or type of surface.

## 7. Conclusions

Once the improvement in the algorithm used was achieved, we delivered more information regarding the possibility of predicting a result. That is why we integrated the probability of victory to our analysis. This probability indicates, given a match and evaluating the ranking difference between the players that intervene in it, with which percentage our model may predict that the best ranked player will win.

For that, and based on a model of logistic regression, we have compared how our model behaved with respect to other existent models. This comparison proved our model more optimistic than the rest. We needed, then, a certainty calculation to see our optimism did not indicate something that was not accurate. By using the AUROC measure we could observe that our model, besides being optimistic, has a high certainty margin.

Given all these results, we can affirm that, throughout this work, we have created a new robust tennis ranking, capable to predict with a high percentage the result of the matches, considering a greater detail, as the ranking difference, to be able to say with what certainty we are going to sense who will be the winner of the encounter.

# Bibliography

[1] James B. *The ranking that change tennis.* From http://www.atpworldtour. com/news

[2] ATP World Tour *Emirates ATP Rankings FAQ.* From http://www. com/news

[3] ATP World Tour *About the ATP Challenger* From http://www. atpworldtour.com/en/corporate/history

[4] Bryant H. *How to fix tennis' big problems* 2013. From http://espn.go.com/ com/tennis

[5] LLangville, Amy N., and Carl D. Meyer. *Who is #1?: the science of rating and ranking.* Princeton University Press, 2012.

[6] Brin, Sergey, and Lawrence Page. *Reprint of: The anatomy of a large-scale hypertextual web search engine.* Computer networks 56.18 (2012): 3825-3833.

[7] Ashish G. *Applications of PageRank to Recommendation Systems.* From http://web.stanford.edu/class/msande233/handouts/lecture8.pdf

[8] Page L., Brin S., Motwani R. and Winograd T *The PageRank Citation Ranking: Bringing Order to the Web* 7th International World Wide Web Conference, Brisbane, Australia, 1998.

[9] Irons, David J., Stephen Buckley, and Tim Paulden. *Developing an improved tennis ranking system.* Journal of Quantitative Analysis in Sports 10.2 (2014): 109-118.

[10] Radicchi, Filippo, and Matjaz Perc. *Who is the best player ever? A complex network analysis of the history of professional tennis.* PloS one 6.2 (2011): e17249.

[11] Dingle, Nicholas, William Knottenbelt, and Demetris Spanias. *On the (page) ranking of professional tennis players.* Computer Performance Engineering. Springer Berlin Heidelberg, 2013. 237-247.

[12] Spanias, A. Demetris, and B. William Knottenbelt. *Tennis Player Ranking using Quantitative Models.* Manuscript

[13] Blackburn, McKinley L. *Ranking the performance of tennis players: an application to womens professional tennis.* Journal of Quantitative Analysis in Sports 9.4 (2013): 367-378.

[14] Barrow, Daniel, et al. *Ranking rankings: an empirical comparison of the predictive power of sports ranking methods.* Journal of Quantitative Analysis in Sports 9.2 (2013): 187-202.

[15] Dormagen, David. *Development of a Simulator for the FIFA World Cup 2014.* Bachelorarbeit FU Berlin 13 (2014). Manuscript

[16] Anagnostopoulos C., Hand D.J., Adams N.M *Measuring classification performance: the hmeasure package* Department of Mathematics, South Kensington Campus, Imperial College London 2012

[17] Fawcett, Tom. *ROC graphs: Notes and practical considerations for researchers.* Machine learning 31 (2004): 1-38.

[18] Leskovec, Jure, Anand Rajaraman, and Jeffrey David Ullman. *Mining of massive datasets.* Princeton University Press, 2014.

# 7. APPENDIX

## 7.1. Database

As we mentioned in the section we spoke about the data set, we have files that contained information of about 40,000 matches.

Once these files where obtained, the task of interpreting the data of the .csv files was not easy. The files had different formats, for some years we had more information and for others we did not have all the required information.

Firstly, we needed to standardize the names of the players, since there were matches where the player was Juan Martin Del Potro and others where the name of the player was J.M Del Potro.

For that we built a script that took care of evaluating players who had one game and evaluate if there was a player with a similar name, or they were players with just one ATP match in those 10 years being evaluated.

The same condition occurred with the types of tournaments and its names, since throughout all of these years the tournaments changed their names. For example, the ATP 250 and ATP 500 both were called International Series. All of this information was added in order to have a reliable data set.

Other of the complications found is that the order of the information provided for every year was different. In some files we had the result in a column while in a different year the information was in other column, that it is why every file was reordered to, by using a script, interpret in a simple way each one of the files.

Once the standardization of all the data was made we divided the information into 3 tables.

In order to do that, by going through each file and obtaining a match per line, we generated three tables that contained all the necessary information.

- Players: Player _ID - Name of the player

- Tournaments: Tournament ID - Name - Year - Week - Surface - Number of sets - Type of tournament - Place - Ceiling

- Matches: Player ID - Winner ID - Loser ID - Ranking ATP Winner - Ranking ATP Loser - WSet1 - LSet1 - WSet2 - LSet2 - WSet3 - LSet3 - WSet4 LSet4 - WSet5 - LSet5 - Winner Sets - Loser Sets - Full Match - Bet Payment Winner - Bet Payment Loser

It was important to evaluate each player to check it was not already on our database and to be able to associate each player of the tournaments to our base. At the same time, we had the exact date of the match in the gathered data, so the script that inserted the tournaments calculated, given a specific date, to what week and year it corresponded.

An extra complication that we had to sort out was that not all the games had with the ATP ranking information, so we generated a script that crawled the official site of the ATP (www.atpworldtour.com) to obtain the corresponding ranking for the players of those matches.

## 7.2. Implementation of the new PageRank model

### 7.2.1. Parametric Pagerank

In order to calculate the parametric PageRank, we have developed the corresponding scrips in Python that will allow us to calculate the PageRank given a specific set of tournaments.

The calculation process was as it follows:

- Graph generation by match

  To generate a ranking for a particular tournament, our script looks at every tournament with a maximum of a certain age. For each one of these tournaments, we look at every match played and traced an edge directed from the loser to the winner. For each edge previously traced, the weight that will be given to that match will be calculated. To do that, as it was explained, it is calculated how much that match sums in terms of age, surface, type, and tournament round, compared to the original tournament for which the ranking will be generated.

  Once the tournament is evaluated, each edge is added to a directed multigraph.

- Obtaining of a PageRank per tournament and ranking generation

  Once the multigraph is obtained with all the games evaluated for the tournament, through the library, the *pagerank_spicy* method is used that will give us a score for each vertex of the multigraph. This score is the one that, according to its results, a player will receive in the given raking.

  When those scores are achieved, each player is linked to a name and the data is ordered in a descending manner using the score. With such order we get an arrangement where the first element is the number one player of the ranking and so we generate a general ranking.

- Evaluation of results based on the obtained ranking

  By knowing how the PageRank is generated, we evaluate how this calculation behaves in each tournament in every year. And that is why that by going through each tournament, the ranking is generated and in each match it is evaluated if the ranking indicated correctly that the winner of the match had a better positioning in the ranking than the loser. In case it is correct, we consider it a *Hit;* in case it is not, we consider it a *Miss.* It should be taken into consideration that we only evaluate the matches that resulted to be complete. Once the hit and miss of the entire year are obtained, we evaluated how our parametric PageRank model behaved for that year.

The implementation for the PageRank calculation allows us, by using parameters, to indicate which attributes are intended to be evaluated for a run. That allow us to differentiate by surface, age, type of tournament, round, among other runs made to improve our model. For those new players in the ranking, it is assigned manually in our PageRank a very high ranking, since it has no history that reflects in the multigraph that is used to calculate the PageRank.

### 7.2.2. Best parameters calculation

To find the best combination of parameters to be used to calculate the value of every edge we used the following algorithm

---

**Algorithm 1** Algorithm browser of new parameters.

*yearDone* ← *true*
*surfaceDone* ← *true*
*tournamentDone* ← *true*
*exponentialDone* ← *true*

*exponential* ← [5,10]
*surfaces* ← [00,1.,1]
*tournaments* ← [11,1.,2]
*years* ← [1.,6]

*maxResult* ← 0
*bestExponentialDecay* ← 0
*bestTournamentWeight* ← 0
*bestYearBefore* ← 0
*bestSurfaceWeight* ← 0

---

**while** *yearDone* and *surfaceDone* and *tournamentDone* and *exponentialDone* **do**
    **for** *y* in *years* **do**
        (*result,year*) ← *max*(*processYear*(*y*))
    **end for**

    **if** *result == maxResult* **then**
        *yearDone* ← *false*
    **end if**

    **if** *result > maxResult* **then**
        *bestYearBefore* ← *year*
        *maxResult* ← *result*
    **end if**

    **for** *e* in *exponential* **do**
        (*result,exponential*) ← *max*(*processExponential*(*e*))
        **end for**

    **if** *result == maxResult* **then**
        *yearDone* ← *false*
        **end if**

    **if** *result > maxResult* **then**
    *bestExponentialDecay* ← *exponential*
        *maxResult* ← *result*
        **end if**

    **for** *s* in *surface* **do**
        (*result,surface*) ← *max*(*processSurface*(*s*))
        **end for**

    **if** *result == maxResult* **then**
    *surfaceDone* ← *false*
        **end if**

    **if** *result > maxResult* **then**
        *bestSurfaceWeight* ← *surface*
        *maxResult* ← *result*
        **end if**

    **for** *t* in *tournaments* **do**
        (*result,tournament*) ← *max*(*processTournament*(*t*))
        **end for**

    **if** *result == maxResult* **then**
    *tournamentDone* ← *false*
        **end if**
    **if** *result > maxResult* **then**
        *bestTournamentWeight* ← *tournament*
        *maxResult* ← *result*
        **end if**
    **end while**

---

**Algorithm 2** Algorithm browser for best *year.*
**Procedure** processYear (y)

*bestYearBefore* ← *y*

*result* ← *evaluateTournamentWithBetterParameters*()

return *result*

**end procedure**

---

**Algorithm 3** Algorithm browser for best *exponential.*
   **procedure** processExponential(*e*)

*bestExponentialDecay* ← *e*

*result* ← *evaluateTournamentWithBetterParameters*()

return *result*

**end procedure**

---

**Algorithm 4** Algorithm browser of the best parameter for surface.

**procedure** processSurface(*s*)

*bestSurfaceWeight* ← *s*

*result* ← *evaluateTournamentWithBetterParameters*()

return *result*

**end procedure**

---

**Algorithm 5** Algorithm browser of the best λ for tournament.

**procedure** processTournament(*t*)

*bestTournamentWeight* ← *t*

*result* ← *evaluateTournamentWithBetterParameters*()

return *result*

**end procedure**

---

Where *evaluateTournamentWithBetterParameters* is the function that calculates the efficiency of the Pagerank with the parameters in the flags as the best parameters of that run.

### 7.2.3. Probability of victory and AUROC

For the calculation of victory and the posterior AUROC analysis, we have generated a table in the database that contains as information: winner ranking, loser ranking, and match number. In order to complete this table, we have evaluated each game stored and saved the ranking that delivered our model as a result for each player.

- Calculation of victory probability

    Evaluating each record in the generated table, we calculated the ranking difference for each game and also we crossed the information with the rest of the tables in order to know the surface and type of tournament being played. Then we group the ranking difference and calculated the prediction efficiency of our ranking for that difference.

That gives us place to generate two vectors. On one hand the *xdata* vector, where we have all the ranking differences, and on the other hand the *ydata* vector, that would have the efficiency of our ranking for those differences.

It is here where we try to minimize the error and generate the curve that manages to draw each of those points closer as we indicated in the "Victory probability" section. For that we calculated the curve that brings the function closer using the *curve fit* method of the *sklearn* library. Once the best parameter for our function of the regression model used is obtained, we are able to evaluate given a difference of the ranking, how optimistic our model is when predicting the result.

Considering that the search could be parametrized, since we had the information of those matches such as surface and type of tournament, we started to generate the same for each possible combination.

That allowed us to generate a decision tree where with the characteristics of a match given, besides the ranking difference, we could provide a greater certainty of the results of the match.

- AUROC calculation

For the calculation of the area under the ROC curve we re ran each game again, and we set the tuples where the first element represented if the match was a *hit* or a *miss* and the second element was the P(victory) of that game.

In the same way the result was replicated in the opposite way; this allow us to generate a type *S* graphic necessary to calculate the ROC curve correctly and thus obtaining the area under the curve.

For the ROC curve calculation, we generated two arrays: *yTrue*, that had the first elements of each tuple, and the *yScore*, that had the second element. Then, with the *roc curve* function, we generated the ROC curve that allow us to indicate the *False Positive Rate and the True Negative Rate.*

In order to obtain the AUROC we used the *roc auc score* method, that with given the generated arrays, already delivered the result of the area.